\documentclass[12pt]{article}

\usepackage{times}
\usepackage{geometry}
\usepackage[utf8]{inputenc}
\usepackage{enumitem,amssymb}
\usepackage{ragged2e}
\newlist{thematic}{itemize}{8}
\setlist[thematic]{label=$\square$}
\usepackage{pifont}
\newcommand{\cmark}{\ding{51}}%
\newcommand{\done}{\rlap{$\square$}{\raisebox{2pt}{\large\hspace{1pt}\cmark}}%
\hspace{-2.5pt}}

\usepackage{graphicx}
\usepackage[numbers]{natbib}
\usepackage{url}
\usepackage{hyperref}
\usepackage{float}
\usepackage{wrapfig}
\usepackage{verbatim}
\usepackage{color}

\begin{document}
\thispagestyle{empty}
\raggedright
\huge
Astro2020 Science White Paper \linebreak

Construction of an $L_*$ Galaxy: the Transformative Power of Wide Fields for Revealing the Past, Present and Future of the Great Andromeda System  \linebreak
\normalsize

\noindent \textbf{Thematic Areas:} \hspace*{60pt} 
\linebreak
  \done Resolved Stellar Populations and their Environments \hspace*{40pt} \linebreak
  \done    Galaxy Evolution   \hspace*{45pt} 
  \linebreak
  
\textbf{Principal Author:}

Name: Karoline M.\ Gilbert	
 \linebreak						
Institution: Space Telescope Science Institute / Johns Hopkins University  
 \linebreak
Email: kgilbert@stsci.edu
 \linebreak
Phone: 1 (410) 338-2475
 \linebreak
 
\textbf{Primary Co-author:}
  \linebreak
  Erik J.\ Tollerud, Space Telescope Science Institute (etollerud@stsci.edu)
  \linebreak
 
\textbf{Co-authors:}
  \linebreak
Jay Anderson, Space Telescope Science Institute
  \linebreak
  Rachael L.\ Beaton, Princeton University
  \linebreak
Eric F.\ Bell, University of Michigan
  \linebreak
Alyson Brooks, Rutgers, the State University of New Jersey
  \linebreak
Thomas M.\ Brown, Space Telescope Science Institute
  \linebreak
James Bullock, UC Irvine
  \linebreak
Jeffrey L.\ Carlin, Large Synoptic Survey Telescope
  \linebreak
Michelle Collins, University of Surrey
  \linebreak
Andrew Cooper, National Tsing Hua University, Taiwan
  \linebreak
Denija Crnojevi\'c, University of Tampa
  \linebreak
Julianne Dalcanton, University of Washington
  \linebreak
Andr\'es del Pino, Space Telescope Science Institute
  \linebreak
Richard D’Souza, University of Michigan
  \linebreak
Ivanna Escala, California Institute of Technology
  \linebreak
Mark Fardal, Space Telescope Science Institute
  \linebreak
Andreea Font, Liverpool John Moores University
  \linebreak
Marla Geha, Yale University
  \linebreak
Puragra Guhathakurta, University of California Santa Cruz
  \linebreak
Evan Kirby, California Institute of Technology
  \linebreak
Geraint F.\ Lewis, The University of Sydney
  \linebreak
Jennifer L.\ Marshall, Texas A\&M University
  \linebreak
Nicolas F.\ Martin, CNRS/INSU, Universit\'e de Strasbourg
  \linebreak
Kristen McQuinn, Rutgers University
  \linebreak
Antonela Monachesi, Universidad de La Serena, Chile
  \linebreak
Ekta Patel, University of Arizona
  \linebreak
Molly S.\ Peeples, Space Telescope Science Institute / Johns Hopkins University
  \linebreak
Annalisa Pillepich, Max Planck Institute for Astronomy, Heidelberg
  \linebreak
Amanda C.\ N.\ Quirk, UC Santa Cruz
  \linebreak
R.\ Michael Rich, University of California, Los Angeles
  \linebreak
S.\ Tony Sohn, Space Telescope Science Institute
  \linebreak
Yuan-Sen Ting, Institute for Advanced Study / Princeton University / Carnegie Observatories
  \linebreak
Roeland P.\ van der Marel, Space Telescope Science Institute
  \linebreak
Andrew Wetzel, University of California, Davis
  \linebreak
Benjamin F.\ Williams, University of Washington
  \linebreak
Jennifer Wojno, Johns Hopkins University
  \linebreak

\textbf{Abstract:} 
The Great Andromeda Galaxy (M31) is the nexus of the near-far galaxy evolution connection and a principal data point for near-field cosmology. Due to its proximity ($\sim$\,780~kpc), M31 can be resolved into individual stars like the Milky Way (MW). Unlike the MW,  
we have the advantage of a global view of M31, enabling M31 to be observed with techniques that also apply to more distant galaxies.  Moreover, recent evidence suggests that M31 may have survived a major merger within the last several Gyr, shaping the morphology of its stellar halo and triggering a starburst, while leaving the stellar disk largely intact.  The MW and M31 thus provide complementary opportunities for in-depth studies of the disks, halos, and satellites of $L_*$ galaxies.  

Our understanding of the M31 system will be transformed in the 2020s if they include wide field facilities for both photometry (HST-like sensitivity and resolution) and spectroscopy (10~m class telescope, $>1$~deg$^2$ field, highly multiplexed, $R\sim$\,3000\,--\,6000). 
We focus here on the power of these facilities to constrain the past, present, and future merger history of M31, via chemo-dynamical analyses and star formation histories of
phase-mixed stars accreted at early times, as well as stars in surviving tidal debris features, M31's extended disk, and intact satellite galaxies that will eventually be tidally incorporated into the halo.  This will yield an unprecedented view of the hierarchical formation of the M31 system and the subhalos that built it into the $L_*$ galaxy we observe today. 

\thispagestyle{empty}

\pagebreak
\setcounter{page}{1} 
\RaggedRight

\subsubsection*{Reconstructing M31's Merger History}
Stellar halos encode a galaxy’s merger history as well as its earliest stages of star formation. $\Lambda$CDM hydrodynamical simulations predict that halos are built from both stars accreted during mergers and those formed in the host. Yet the relative contributions of these formation avenues remain observationally unconstrained, as do the luminosity function and time of infall of accreted satellite galaxies. This motivates efforts to constrain them in M31.

\noindent
{\bf Measuring M31's Recent Accretion History} A wealth of structures have been discovered throughout the M31 system, primarily through star count maps derived from contiguous imaging \citep[summarized by][]{mcconnachie2018}.  These photometric observations, 
reaching $\sim$\,3 magnitudes below the tip of the red giant branch (RGB) and covering $>400$\,deg$^2$, provide lower limits for the luminosity of recently accreted satellites, via estimates of the luminosity of individual substructures.  Tentative connections between substructures have also been surmised: \citet{mcconnachie2018} estimate that at least five accretion events in the last 3\,--\,4 Gyr are required to produce the 13 most distinctive substructures.
Meanwhile, comparisons with simulations have suggested that a recent M31 accretion event may have been a massive merger, generating M31's largest substructure [the Giant Stellar Stream (GSS)], along with additional substructure in M31's inner regions. 

\begin{figure}[hbt]
    \centering
    \includegraphics[width=0.95\textwidth]{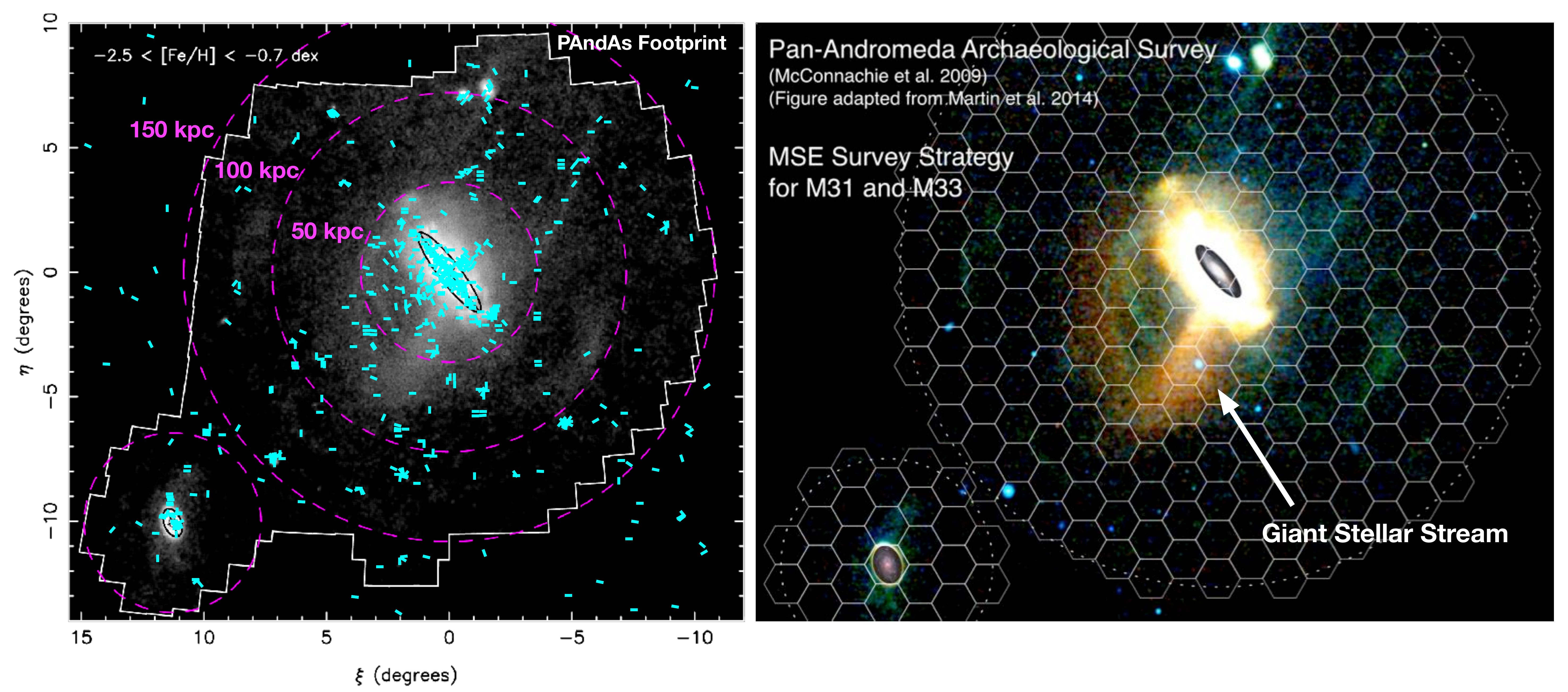}
    \caption{
    {\bf Left panel:} Surface density of red giant branch (RGB) stars in the M31 system \citep[from PAndAS; Fig.\,11 from][]{mcconnachie2018}.   
    Overlaid are all spectroscopic slitmasks observed to date with the DEIMOS spectrograph, which required over 120 nights of observing on the Keck~II telescope, while still covering less than 2.5\% of the area within 150~kpc. 
    {\bf Right panel:} Example tiling strategy for an M31 survey with the proposed Mauna Kea Spectroscopic Explorer (MSE), covering a contiguous area of the halos of M31 and M33 
    out to $\sim$ one-half and one-third 
    the virial radii.  MSE will have the wide field of view (1.5~deg$^2$), high multiplexing (several thousand fibers), 10~m class mirror, and high throughput needed for chemo-dynamical studies of RGB stars in M31's stellar halo. 
    A magnitude-limited, spectroscopic census of every RGB candidate in the outer regions ($\sim$\,40\,--\,150~kpc; 350~deg$^2$) of M31's halo would require $\sim$\,150~nights, 
    while a census of the high density inner regions (40~deg$^2$), with an abundance of bright targets, would only require $\sim$\,15 nights 
    (MSE Science Reference Observations).
    Given a telescope dedicated to spectroscopic surveys, this could be reasonably achieved within a survey duration of 5\,--\,10 calendar years, comparable to the duration over which the contiguous photometric imaging was obtained.  
    }
    \label{fig:m31m33_roadmap}
\end{figure}

Line-of-sight velocities provide a crucial additional observational dimension for determining
connections between substructures and constraining orbits of progenitors. 
However, the small field of view ($\sim$\,0.02~deg$^2$) 
of the best available instrumentation has limited such studies to pencil beam surveys (Fig.\,\ref{fig:m31m33_roadmap}), resulting in only a few hundred stars observed in 
the GSS and its associated debris, and tens of stars or fewer in fainter features.  Thus,  connections between most of M31's substructures remain ambiguous.    

\begin{figure}[hbt]
    \centering
    \includegraphics[width=0.95\textwidth]{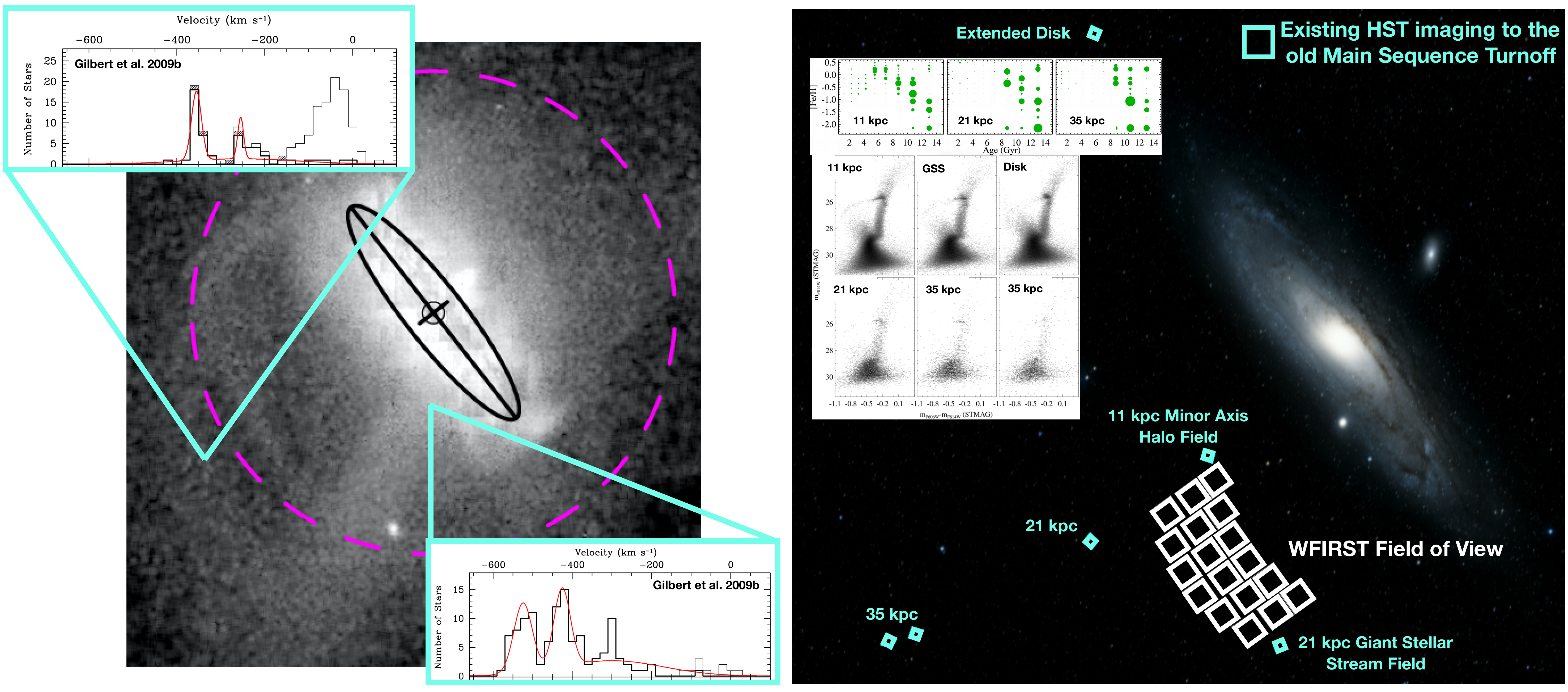}
    \caption{{\bf Left panel:} Spectroscopic observations 
    provide line of sight stellar velocity distributions, 
    providing crucial phase-space constraints for modelling the interactions that produce tidal debris features.  Moreover, they can reveal faint substructures not identified in resolved stellar imaging and 
    resolve spatially overlapping streams. 
    {\bf Right panel:} Locations of all Hubble imaging to the depth of the old main sequence turnoff in M31, with associated color-magnitude diagrams and star formation histories \citep[showing the relative weighting of the stellar population in metallicity and age;][]{brown2008,brown2009}.  M31's stellar halo is not uniformly old, 
    with a significant population of intermediate-age stars in some sight lines.  The fidelity of the SFHs at the largest halo distances are limited by the number of stars available in HST's small field of view.  WFIRST's much larger field of view will enable analogous, spatially contiguous star formation history studies to be performed over a significant portion of M31's inner halo.  
    }
    \label{fig:obs}
\end{figure}

{\it A wide field of view ($\gtrsim$\,1~deg$^2$), highly multiplexed (several thousand targets), medium resolution ($\sim$\,3000\,--\,6000) spectrograph on an 8\,--\,10~m class telescope in the northern hemisphere will fundamentally change our understanding of M31's stellar halo and the substructures within.}  Such a facility could perform a contiguous, magnitude-limited, {\it spectroscopic} mapping of M31's stellar halo.  
Line-of-sight kinematics along the full length of tidal debris features will 
differentiate related and unrelated structures, 
reveal faint, undetected substructure \citep[e.g.,][]{gilbert2007,gilbert2012} and resolve spatially overlapping streams indistinguishable in star count maps \citep[Fig.\,\ref{fig:obs}; e.g.,][]{ibata2005,gilbert2009gss}.

Mean proper motions and precision distances of individual substructures will also be measureable in the 2020s 
(with HST, JWST and WFIRST). 
Combined with spatially resolved line of sight velocities and dispersions, 
and contiguous resolved star count maps, this will yield stringent 6D phase-space constraints 
on orbital modelling.  
This will provide an accurate count of the number of unique surviving substructures in the stellar halo of M31 (and M33), yielding true estimates, rather than lower limits, of the number, luminosities, and time of first disruptive pericentric passage for M31 satellites destroyed within the last several Gyr.  

\noindent
{\bf Ancient Star Formation Histories of Destroyed Satellites}
Only 10\% of the stellar mass in M31's stellar halo (27$<R<$150~kpc) is in the tidal debris features discussed above: 86\% of the stellar mass is in either amorphous substructure (59\%) or a  ``smooth'' component (27\%), with the remaining 4\% in surviving satellites \citep{mcconnachie2018}.  Global properties of stellar halos in cosmological simulations have been shown to quantitatively correlate with properties such as the halo formation time, the total mass of accreted stars, and the time since the last major merger \citep[e.g.,][]{pillepich2014}.  
This has enabled inferences regarding M31's ancient accretion history based on observables such as M31's halo surface brightness and metallicity profiles \citep[e.g.,][]{gilbert2012,gilbert2014}. 
Furthermore, the latest generation of hydrodynamical simulations can follow the chemical enrichment of halo progenitors down to masses as low as $10^6 M_\odot$ for hundreds of M31 analogues, providing opportunities to quantitatively compare M31's stellar halo with suites of simulated accretion histories. However, we 
lack the data to 
confirm
any hypotheses derived from such comparisons. 

Transformational progress in determining the origin of the majority of M31's stellar halo mass can be made in the next decade with the combination of large-scale chemical abundance measurements and imaging of stars below the old main sequence turnoff (oMSTO). Both provide estimates of the ancient star formation history (SFH).  
High-metallicity ([Fe/H]) stars form in environments  
with significant past star formation, such as massive, efficiently star-forming satellites. Stars with high [$\alpha$/Fe] ratios form in regions enriched by massive, $\alpha$-rich supernovae, but not yet enriched by low-mass, Fe-rich supernovae. Satellites that fell in and quenched early on will have truncated SFHs and high [$\alpha$/Fe]. Resolved stellar imaging below the oMSTO provides a direct measure of the age and metallicity distribution of stars.  This has been achieved with Hubble for a handful of isolated lines of sight in M31's stellar halo. WFIRST will provide similar resolution and sensitivity as HST, but with a field of view nearly 100 times larger, enabling oMSTO studies of M31's stellar halo over regions large enough to cover and bridge tidal features (Fig.\,\ref{fig:obs}).

\begin{figure}
    \centering
    \includegraphics[width=0.75\textwidth]{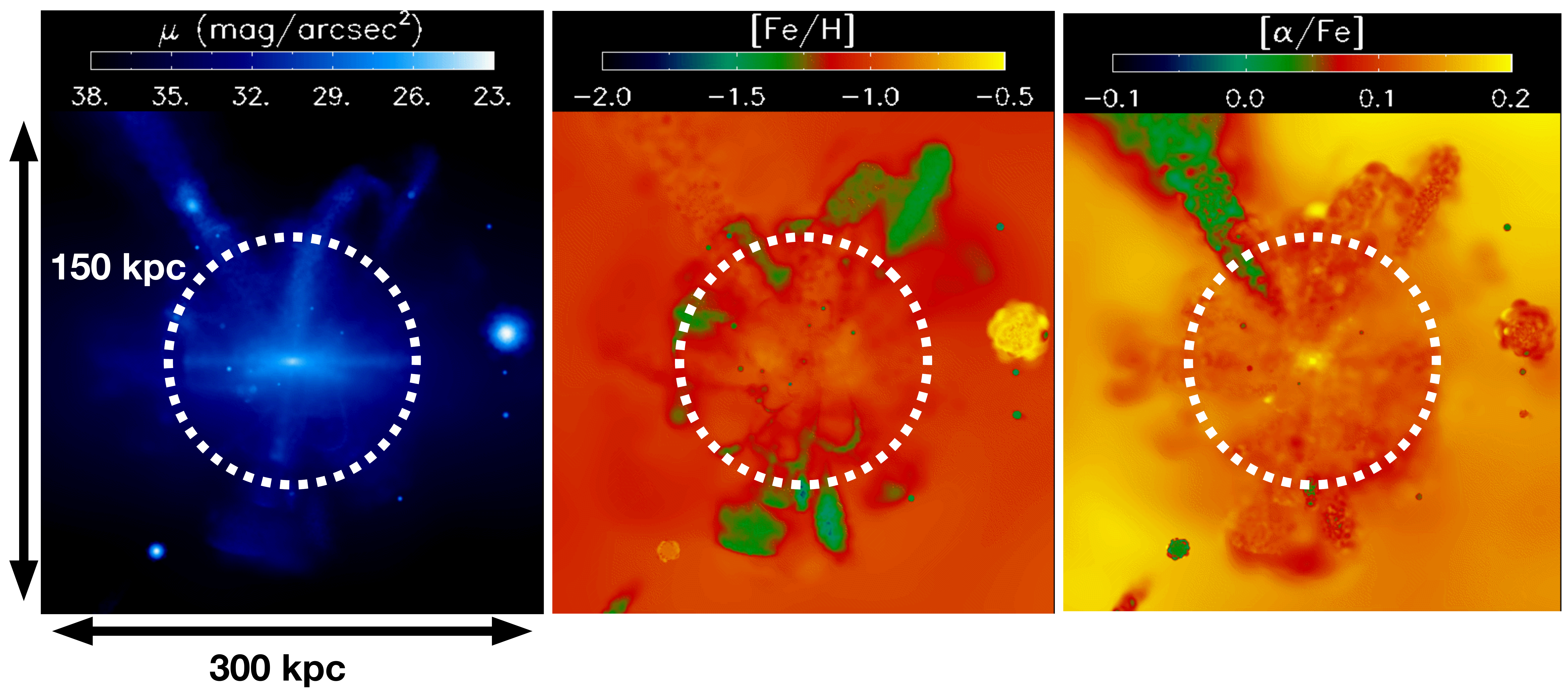}
    \caption{Simulations of stellar halo formation in a cosmological framework make predictions for the spatial distribution of observables such as stellar density and chemical abundances.  These are sensitive to the orbits, luminosity function, and time of accretion of infalling satellites, as well as the fraction of halo stars formed within the host potential.  With the spectroscopic facilities proposed for the 2020s it will be possible
    to produce similar {\it observational} maps for the M31 system, to a significant fraction of the virial radius. 
    Figures depict a \citet{bullock2005} simulated halo, built from accretion \citep[image credit: Sanjib Sharma;][]{gilbert2009a}.  
    }
    \label{fig:sims}
\end{figure}

Current facilities limit direct measurements of [Fe/H] and [$\alpha$/Fe] to a few 
isolated sight-lines 
\citep{vargas2014,vargas2014apjl,escala2019}.  
with R$\sim$\,3000\--\,6000 spectra covering optical to near-infrared wavelengths, but require significantly higher signal to noise ratios than velocity measurements. 
{\it A contiguous, global mapping of [Fe/H] and [$\alpha$/Fe] in M31's stellar halo, achievable with a high throughput, 10~m class spectroscopic survey facility, will yield a statistical measure of the full luminosity distribution of disrupted satellites and their times of infall into M31.}  
In substructures from recent mergers, abundance gradients can be used to characterize the internal structure of the progenitor \citep[e.g.,][]{fardal2008}.  Applied to the inner regions of M31's halo and disk, [Fe/H] and [$\alpha$/Fe] abundances will also constrain the fraction of halo stars formed {\it in situ} in M31's potential \citep[for example in M31's proto-disk; e.g.,][]{zolotov2010,tissera2013}.  Combined, these measurements will enable the first global, quantitative comparisons to detailed galaxy evolution simulations (Fig.\,\ref{fig:sims}).

\noindent
{\bf Has M31's disk survived a major merger?}
In $\Lambda$CDM galaxy formation models, the merging of lower-mass subhalos with larger halos is ubiquitous, frequent, and a primary driver of galaxy evolution. 
A challenge 
of this paradigm is understanding how galaxy disks survive 
\citep{toth1992}. 

M31's largest tidal debris feature, the GSS, has been reproduced in detail by numerical simulations of a minor merger \citep[with an $\sim$\,LMC mass system; e.g.,][]{fardal2013}.  However, large-volume hydrodynamical simulations now resolve large numbers of M31 analogues, providing a statistical sample of systems
with and without a recent major merger event.  Recent work based on these simulations has argued that a wide range of observations in M31 (the GSS and assocated debris, disturbed disk morphology, a disk-wide burst of star formation 2\,--\,4~Gyr ago, a 
disk population with halo-like kinematics, and more) might be explained by a {\it major} merger occurring within the last several Gyr \citep{hammer2018,dsouza2018}.  {\it If true, this would provide an invaluable local case study 
of the effect of major mergers on star formation, galaxy structure, and disk survival in $L_*$ galaxies.}

Unfortunately, the parameter space for simulating major merger scenarios detailed enough to be compared to actual M31 observations, 
and to confirm a massive GSS progenitor, is intractably large. Significant progress in differentiating between a major or minor merger will only be made by increasing the observational constraints.  
A 400-pointing HST program \citep[PHAT;][]{dalcanton2012} established the universal nature of the 2\,--\,4~Gyr SF burst in the northeast half of the disk \citep{williams2017} and identified isolated targets for ground-based 
spectroscopy, which measured kinematical properties of the disk and halo that could be signs of a recent major merger \citep{dorman2013,dorman2015}. A minimal investment with WFIRST ($\sim$\,2 pointings for an equivalent footprint in the southwest) will provide high resolution imaging over the remainder of M31's star forming disk, providing contiguous SFH maps and isolated spectroscopic targets throughout the side of the disk closest to a proposed remnant of the major merger progenitor \citep[M32;][]{dsouza2018}.

Existing spectroscopic facilities can efficiently survey the inner disk regions.  However, the sparser outer regions of M31's disk and inner halo is where disturbances and debris that are likely to be most useful in distinguishing between the minor or major merger scenario will reside.  This area is too large to contiguously survey with present facilities. {\it A dedicated spectroscopic survey facility 
would efficiently tile this region (Fig.\,\ref{fig:m31m33_roadmap}), producing an unprecedented map of stellar kinematics and chemical abundances throughout the extended disk and inner halo of M31.}

\subsubsection*{Dwarf Galaxies and the edge of M31's Halo}

While the stellar halo of M31 illuminates M31's past, M31's surviving satellite galaxies are a view into its future.
They are interesting and unique in their own right: And II is a possible merger remnant \citep{ho12}, M32 is a cE \citep{dsouza2018}, M33 is expected to host satellites of its own \citep{patel2018}, and either a major merger or LMC-mass minor merger would likely have contributed satellites to the M31 halo, which remain unidentified. 
Moreover, M31's dwarf satellites provide a crucial 
comparison to 
MW satellites.
For example, the luminosity function of M31 satellites reveals a  higher normalization for M31, hinting at a higher halo mass 
\citep{yniguez14}. 
The M31 satellite population also displays cosmological conundrums like the MW's, including the too big to fail and core-cusp ``problems'' 
\citep[][]{tollerud14, collins14} and the plane of co-rotating satellites \citep{ibata13planes}. 
M31 provides the first (and often only) cross-check on the universal applicability of arguments centered around the MW satellite system.

However, current spectroscopy of these satellites is highly non-uniform, ranging from 1 to over 20 hours invested per dwarf. 
Significant improvements in our understanding of the M31 satellite system 
will require homogeneous stellar spectroscopy of the M31 faint satellite population, particularly in their outskirts where tidal effects are apparent.  This will enable statistical comparisons to simulations, the MW, and most importantly, external galaxies where studies now only possible in the LG will be feasible in the 2020s.  Fortunately, such a survey comes ``for free'' as part of an M31 stellar halo survey like that outlined above.  The only added requirement would be a compact ($\sim$\,few arcmin) area of relatively high fiber density ($\sim$\,10 stars per $100 \; {\rm arcsec}^2$).  

Moreover, a truly transformative expansion of our understanding of the M31 satellite system and halo would be possible with a highly multiplexed spectroscopic survey extending to M31's virial radius ($\sim$\,300~kpc). 
Such a survey would find M31 halo RGBs through raw numbers: experience with Keck/DEIMOS and extrapolation of the known halo profile \citep{gilbert2018} suggests that a survey spectrograph capable of $\sim$\,4000 spectra per ${\rm deg}^2$ would yield tens to hundreds of halo stars at 300 kpc. 
As shown in Fig.~3, the outer halo often contains qualitatively different, less phase-mixed features. 
Such a survey would result in the only $\sim L_*$ halo beyond the MW with a census of halo tracers (satellites, clusters, and stars) to the virial radius.  It would provide the last necessary element (with HST, JWST, and WFIRST proper motions) for complete 6D datasets throughout M31's halo in the 2020s, providing 
unique 
insights into M31's dark matter halo profile.

\subsubsection*{Expanding Resolved Stellar Halo Studies Beyond the Local Group}
The facilities and surveys of the 2020s will set the stage for detailed tests of the $\Lambda$CDM paradigm in galaxies beyond the Local Group, covering for the first time a significant range of galactic mass, morphology, and environment.  Contiguous, resolved star imaging of the RGB in halos of more distant galaxies will be obtained with the wide field of view and high spatial resolution of WFIRST ($\lesssim 10$~Mpc).  Multi-object spectrographs on 30-m class telescopes will enable spectroscopic studies of small regions of these halos, targeting the brightest tidal features ($\lesssim 3.5$~Mpc).  
JWST will provide oMSTO imaging, 
constraining ancient SFHs in galaxies to $\sim$\,3~Mpc (in $\lesssim$\,100~hr),
while a space-based 15~m (9~m)  UV-optical-NIR telescope could extend studies like PHAT to galaxies as far as $\sim$\,7 (4) Mpc 
and ancient SFH studies to $\sim$\,10 (5)~Mpc. 
In short, within the Local Volume we will obtain observations like those currently available for the M31 system. 
The understanding of M31's merger history produced by the surveys described above will be the foundation on which we will interpret this new observational frontier.  

\bibliographystyle{yahapj}
\bibliography{main}

\begin{thebibliography}{}
\providecommand\natexlab[1]{#1}
\providecommand\JournalTitle[1]{#1}

\bibitem[{{Brown} {et~al.}(2008){Brown}, {Beaton}, {Chiba}, {Ferguson},
  {Gilbert}, {Guhathakurta}, {Iye}, {Kalirai}, {Koch}, {Komiyama}, {Majewski},
  {Reitzel}, {Renzini}, {Rich}, {Smith}, {Sweigart}, \& {Tanaka}}]{brown2008}
{Brown}, T.~M., {Beaton}, R., {Chiba}, M., {et~al.} 2008,
  \href{http://dx.doi.org/10.1086/592686}{\JournalTitle{\apjl}, 685, L121}

\bibitem[{{Brown} {et~al.}(2009){Brown}, {Smith}, {Ferguson}, {Guhathakurta},
  {Kalirai}, {Kimble}, {Renzini}, {Rich}, {Sweigart}, \& {Vanden
  Berg}}]{brown2009}
{Brown}, T.~M., {Smith}, E., {Ferguson}, H.~C., {et~al.} 2009,
  \href{http://dx.doi.org/10.1088/0067-0049/184/1/152}{\JournalTitle{\apjs},
  184, 152}

\bibitem[{{Bullock} \& {Johnston}(2005)}]{bullock2005}
{Bullock}, J.~S., \& {Johnston}, K.~V. 2005,
  \href{http://dx.doi.org/10.1086/497422}{\JournalTitle{\apj}, 635, 931}

\bibitem[{{Collins} {et~al.}(2014){Collins}, {Chapman}, {Rich}, {Ibata},
  {Martin}, {Irwin}, {Bate}, {Lewis}, {Pe{\~n}arrubia}, {Arimoto}, {Casey},
  {Ferguson}, {Koch}, {McConnachie}, \& {Tanvir}}]{collins14}
{Collins}, M. L.~M., {Chapman}, S.~C., {Rich}, R.~M., {et~al.} 2014,
  \href{http://dx.doi.org/10.1088/0004-637X/783/1/7}{\JournalTitle{\apj}, 783,
  7}

\bibitem[{{Dalcanton} {et~al.}(2012){Dalcanton}, {Williams}, {Lang}, {Lauer},
  {Kalirai}, {Seth}, {Dolphin}, {Rosenfield}, {Weisz}, {Bell}, {Bianchi},
  {Boyer}, {Caldwell}, {Dong}, {Dorman}, {Gilbert}, {Girardi}, {Gogarten},
  {Gordon}, {Guhathakurta}, {Hodge}, {Holtzman}, {Johnson}, {Larsen}, {Lewis},
  {Melbourne}, {Olsen}, {Rix}, {Rosema}, {Saha}, {Sarajedini}, {Skillman}, \&
  {Stanek}}]{dalcanton2012}
{Dalcanton}, J.~J., {Williams}, B.~F., {Lang}, D., {et~al.} 2012,
  \href{http://dx.doi.org/10.1088/0067-0049/200/2/18}{\JournalTitle{\apjs},
  200, 18}

\bibitem[{{Dorman} {et~al.}(2013){Dorman}, {Widrow}, {Guhathakurta}, {Seth},
  {Foreman-Mackey}, {Bell}, {Dalcanton}, {Gilbert}, {Skillman}, \&
  {Williams}}]{dorman2013}
{Dorman}, C.~E., {Widrow}, L.~M., {Guhathakurta}, P., {et~al.} 2013,
  \href{http://dx.doi.org/10.1088/0004-637X/779/2/103}{\JournalTitle{\apj},
  779, 103}

\bibitem[{{Dorman} {et~al.}(2015){Dorman}, {Guhathakurta}, {Seth}, {Weisz},
  {Bell}, {Dalcanton}, {Gilbert}, {Hamren}, {Lewis}, {Skillman}, {Toloba}, \&
  {Williams}}]{dorman2015}
{Dorman}, C.~E., {Guhathakurta}, P., {Seth}, A.~C., {et~al.} 2015,
  \href{http://dx.doi.org/10.1088/0004-637X/803/1/24}{\JournalTitle{\apj}, 803,
  24}

\bibitem[{{D'Souza} \& {Bell}(2018)}]{dsouza2018}
{D'Souza}, R., \& {Bell}, E.~F. 2018,
  \href{http://dx.doi.org/10.1038/s41550-018-0533-x}{\JournalTitle{Nature
  Astronomy}, 2, 737}

\bibitem[{{Escala} {et~al.}(2018){Escala}, {Kirby}, {Gilbert}, {Cunningham}, \&
  {Wojno}}]{escala2019}
{Escala}, I., {Kirby}, E.~N., {Gilbert}, K.~M., {Cunningham}, E.~C., \&
  {Wojno}, J. 2018, \JournalTitle{ArXiv e-prints},
  \href{http://arxiv.org/abs/1811.09279}{{\sffamily arXiv:1811.09279}}

\bibitem[{{Fardal} {et~al.}(2008){Fardal}, {Babul}, {Guhathakurta}, {Gilbert},
  \& {Dodge}}]{fardal2008}
{Fardal}, M.~A., {Babul}, A., {Guhathakurta}, P., {Gilbert}, K.~M., \& {Dodge},
  C. 2008, \href{http://dx.doi.org/10.1086/590386}{\JournalTitle{\apjl}, 682,
  L33}

\bibitem[{{Fardal} {et~al.}(2013){Fardal}, {Weinberg}, {Babul}, {Irwin},
  {Guhathakurta}, {Gilbert}, {Ferguson}, {Ibata}, {Lewis}, {Tanvir}, \&
  {Huxor}}]{fardal2013}
{Fardal}, M.~A., {Weinberg}, M.~D., {Babul}, A., {et~al.} 2013,
  \href{http://dx.doi.org/10.1093/mnras/stt1121}{\JournalTitle{\mnras}, 434,
  2779}

\bibitem[{{Gilbert} {et~al.}(2009{\natexlab{a}}){Gilbert}, {Font}, {Johnston},
  \& {Guhathakurta}}]{gilbert2009a}
{Gilbert}, K.~M., {Font}, A.~S., {Johnston}, K.~V., \& {Guhathakurta}, P.
  2009{\natexlab{a}},
  \href{http://dx.doi.org/10.1088/0004-637X/701/1/776}{\JournalTitle{\apj},
  701, 776}

\bibitem[{{Gilbert} {et~al.}(2007){Gilbert}, {Fardal}, {Kalirai},
  {Guhathakurta}, {Geha}, {Isler}, {Majewski}, {Ostheimer}, {Patterson},
  {Reitzel}, {Kirby}, \& {Cooper}}]{gilbert2007}
{Gilbert}, K.~M., {Fardal}, M., {Kalirai}, J.~S., {et~al.} 2007,
  \href{http://dx.doi.org/10.1086/521094}{\JournalTitle{\apj}, 668, 245}

\bibitem[{{Gilbert} {et~al.}(2009{\natexlab{b}}){Gilbert}, {Guhathakurta},
  {Kollipara}, {Beaton}, {Geha}, {Kalirai}, {Kirby}, {Majewski}, \&
  {Patterson}}]{gilbert2009gss}
{Gilbert}, K.~M., {Guhathakurta}, P., {Kollipara}, P., {et~al.}
  2009{\natexlab{b}},
  \href{http://dx.doi.org/10.1088/0004-637X/705/2/1275}{\JournalTitle{\apj},
  705, 1275}

\bibitem[{{Gilbert} {et~al.}(2012){Gilbert}, {Guhathakurta}, {Beaton},
  {Bullock}, {Geha}, {Kalirai}, {Kirby}, {Majewski}, {Ostheimer}, {Patterson},
  {Tollerud}, {Tanaka}, \& {Chiba}}]{gilbert2012}
{Gilbert}, K.~M., {Guhathakurta}, P., {Beaton}, R.~L., {et~al.} 2012,
  \href{http://dx.doi.org/10.1088/0004-637X/760/1/76}{\JournalTitle{\apj}, 760,
  76}

\bibitem[{{Gilbert} {et~al.}(2014){Gilbert}, {Kalirai}, {Guhathakurta},
  {Beaton}, {Geha}, {Kirby}, {Majewski}, {Patterson}, {Tollerud}, {Bullock},
  {Tanaka}, \& {Chiba}}]{gilbert2014}
{Gilbert}, K.~M., {Kalirai}, J.~S., {Guhathakurta}, P., {et~al.} 2014,
  \href{http://dx.doi.org/10.1088/0004-637X/796/2/76}{\JournalTitle{\apj}, 796,
  76}

\bibitem[{{Gilbert} {et~al.}(2018){Gilbert}, {Tollerud}, {Beaton},
  {Guhathakurta}, {Bullock}, {Chiba}, {Kalirai}, {Kirby}, {Majewski}, \&
  {Tanaka}}]{gilbert2018}
{Gilbert}, K.~M., {Tollerud}, E., {Beaton}, R.~L., {et~al.} 2018,
  \href{http://dx.doi.org/10.3847/1538-4357/aa9f26}{\JournalTitle{\apj}, 852,
  128}

\bibitem[{{Hammer} {et~al.}(2018){Hammer}, {Yang}, {Wang}, {Ibata}, {Flores},
  \& {Puech}}]{hammer2018}
{Hammer}, F., {Yang}, Y.~B., {Wang}, J.~L., {et~al.} 2018,
  \href{http://dx.doi.org/10.1093/mnras/stx3343}{\JournalTitle{\mnras}, 475,
  2754}

\bibitem[{{Ho} {et~al.}(2012){Ho}, {Geha}, {Munoz}, {Guhathakurta}, {Kalirai},
  {Gilbert}, {Tollerud}, {Bullock}, {Beaton}, \& {Majewski}}]{ho12}
{Ho}, N., {Geha}, M., {Munoz}, R.~R., {et~al.} 2012,
  \href{http://dx.doi.org/10.1088/0004-637X/758/2/124}{\JournalTitle{\apj},
  758, 124}

\bibitem[{{Ibata} {et~al.}(2005){Ibata}, {Chapman}, {Ferguson}, {Lewis},
  {Irwin}, \& {Tanvir}}]{ibata2005}
{Ibata}, R., {Chapman}, S., {Ferguson}, A.~M.~N., {et~al.} 2005,
  \href{http://dx.doi.org/10.1086/491727}{\JournalTitle{\apj}, 634, 287}

\bibitem[{{Ibata} {et~al.}(2013){Ibata}, {Lewis}, {Conn}, {Irwin},
  {McConnachie}, {Chapman}, {Collins}, {Fardal}, {Ferguson}, {Ibata}, {Mackey},
  {Martin}, {Navarro}, {Rich}, {Valls-Gabaud}, \& {Widrow}}]{ibata13planes}
{Ibata}, R.~A., {Lewis}, G.~F., {Conn}, A.~R., {et~al.} 2013,
  \href{http://dx.doi.org/10.1038/nature11717}{\JournalTitle{\nat}, 493, 62}

\bibitem[{{McConnachie} {et~al.}(2018){McConnachie}, {Ibata}, {Martin},
  {Ferguson}, {Collins}, {Gwyn}, {Irwin}, {Lewis}, {Mackey}, {Davidge},
  {Arias}, {Conn}, {C{\^o}t{\'e}}, {Crnojevic}, {Huxor}, {Penarrubia},
  {Spengler}, {Tanvir}, {Valls-Gabaud}, {Babul}, {Barmby}, {Bate}, {Bernard},
  {Chapman}, {Dotter}, {Harris}, {McMonigal}, {Navarro}, {Puzia}, {Rich},
  {Thomas}, \& {Widrow}}]{mcconnachie2018}
{McConnachie}, A.~W., {Ibata}, R., {Martin}, N., {et~al.} 2018,
  \href{http://dx.doi.org/10.3847/1538-4357/aae8e7}{\JournalTitle{\apj}, 868,
  55}

\bibitem[{{Patel} {et~al.}(2018){Patel}, {Carlin}, {Tollerud}, {Collins}, \&
  {Dooley}}]{patel2018}
{Patel}, E., {Carlin}, J.~L., {Tollerud}, E.~J., {Collins}, M.~L.~M., \&
  {Dooley}, G.~A. 2018,
  \href{http://dx.doi.org/10.1093/mnras/sty1946}{\JournalTitle{\mnras}, 480,
  1883}

\bibitem[{{Pillepich} {et~al.}(2014){Pillepich}, {Vogelsberger}, {Deason},
  {Rodriguez-Gomez}, {Genel}, {Nelson}, {Torrey}, {Sales}, {Marinacci},
  {Springel}, {Sijacki}, \& {Hernquist}}]{pillepich2014}
{Pillepich}, A., {Vogelsberger}, M., {Deason}, A., {et~al.} 2014,
  \href{http://dx.doi.org/10.1093/mnras/stu1408}{\JournalTitle{\mnras}, 444,
  237}

\bibitem[{{Tissera} {et~al.}(2013){Tissera}, {Scannapieco}, {Beers}, \&
  {Carollo}}]{tissera2013}
{Tissera}, P.~B., {Scannapieco}, C., {Beers}, T.~C., \& {Carollo}, D. 2013,
  \href{http://dx.doi.org/10.1093/mnras/stt691}{\JournalTitle{\mnras}, 432,
  3391}

\bibitem[{{Tollerud} {et~al.}(2014){Tollerud}, {Boylan-Kolchin}, \&
  {Bullock}}]{tollerud14}
{Tollerud}, E.~J., {Boylan-Kolchin}, M., \& {Bullock}, J.~S. 2014,
  \href{http://dx.doi.org/10.1093/mnras/stu474}{\JournalTitle{\mnras}, 440,
  3511}

\bibitem[{{Toth} \& {Ostriker}(1992)}]{toth1992}
{Toth}, G., \& {Ostriker}, J.~P. 1992,
  \href{http://dx.doi.org/10.1086/171185}{\JournalTitle{\apj}, 389, 5}

\bibitem[{{Vargas} {et~al.}(2014{\natexlab{a}}){Vargas}, {Geha}, \&
  {Tollerud}}]{vargas2014}
{Vargas}, L.~C., {Geha}, M.~C., \& {Tollerud}, E.~J. 2014{\natexlab{a}},
  \href{http://dx.doi.org/10.1088/0004-637X/790/1/73}{\JournalTitle{\apj}, 790,
  73}

\bibitem[{{Vargas} {et~al.}(2014{\natexlab{b}}){Vargas}, {Gilbert}, {Geha},
  {Tollerud}, {Kirby}, \& {Guhathakurta}}]{vargas2014apjl}
{Vargas}, L.~C., {Gilbert}, K.~M., {Geha}, M., {et~al.} 2014{\natexlab{b}},
  \href{http://dx.doi.org/10.1088/2041-8205/797/1/L2}{\JournalTitle{\apjl},
  797, L2}

\bibitem[{{Williams} {et~al.}(2017){Williams}, {Dolphin}, {Dalcanton}, {Weisz},
  {Bell}, {Lewis}, {Rosenfield}, {Choi}, {Skillman}, \&
  {Monachesi}}]{williams2017}
{Williams}, B.~F., {Dolphin}, A.~E., {Dalcanton}, J.~J., {et~al.} 2017,
  \href{http://dx.doi.org/10.3847/1538-4357/aa862a}{\JournalTitle{\apj}, 846,
  145}

\bibitem[{{Yniguez} {et~al.}(2014){Yniguez}, {Garrison-Kimmel},
  {Boylan-Kolchin}, \& {Bullock}}]{yniguez14}
{Yniguez}, B., {Garrison-Kimmel}, S., {Boylan-Kolchin}, M., \& {Bullock}, J.~S.
  2014, \href{http://dx.doi.org/10.1093/mnras/stt2058}{\JournalTitle{\mnras},
  439, 73}

\bibitem[{{Zolotov} {et~al.}(2010){Zolotov}, {Willman}, {Brooks}, {Governato},
  {Hogg}, {Shen}, \& {Wadsley}}]{zolotov2010}
{Zolotov}, A., {Willman}, B., {Brooks}, A.~M., {et~al.} 2010,
  \href{http://dx.doi.org/10.1088/0004-637X/721/1/738}{\JournalTitle{\apj},
  721, 738}

\end{thebibliography}

\end{document}